\documentstyle [11pt, epsfig]{article}
\begin{document}
 \def\la{\langle}
 \def\ra{\rangle}
\hfill {WM-04-108}

\hfill {\today}

\vskip 1in   \baselineskip 24pt

{
\Large
   \bigskip
   \centerline{FCNC and Rare B Decays in 3-3-1 Models}
 }
\def\bar{\overline}

\centerline{J.-Alexis Rodriguez\footnote{Email: 
jarodriguezl@unal.edu.co}}
\centerline {\it Departmento de Fisica}
\centerline {\it Universidad Nacional de Colombia, Bogota, Colombia}
\vskip .2cm
\centerline{and}
\vskip .2cm
\centerline{Marc Sher\footnote{Email: sher@physics.wm.edu}}
\centerline {\it Particle Theory Group}
\centerline {\it Department of Physics}
\centerline {\it College of William and Mary, Williamsburg, VA 23187, 
USA}
\bigskip

{\narrower\narrower  An interesting extension of the 
Standard Model is based on the electroweak gauge group 
$SU(3)_{L}\times U(1)$.   It requires three generations to cancel 
anomalies, treats the third generation differently than the first 
two, and has a rich phenomenology.  There are several models, 
distinguished by the embedding of the charge operator into the 
$SU(3)_{L}$ group and by the choice of fermion representations.  In 
this Brief Report, we 
consider flavor-changing neutral currents in these models, 
concentrating on the $P-\overline{P}$ mass difference, where 
$P=(K, D, B, B_{s})$, as well as $B\rightarrow 
Kl^{+}l^{-}$, $B\rightarrow\mu^{+}\mu^{-}$ and $B_{s}\rightarrow 
\mu^{+}\mu^{-}$ decays.  Although the $P-\overline{P}$ mass difference
has been considered 
previously in some models, the rare B decays are new.  We find that the strongest 
bounds come from  the $B-\overline{B}$ 
and $B_{s}-\overline{B}_{s}$ mass difference.}
\newpage
\section{Introduction}\label{sec:intro}

One of the more intriguing extensions of the standard model is based 
on the gauge group $SU(3)_{c}\times SU(3)_{L}\times U(1)$.   In the 
original, minimal version of the model\cite{frampton,pisano}, the 
charged leptons and  neutrinos  are put into 
antitriplets of $SU(3)_{L}$, two generations of left-handed quarks are 
put into triplets and the other generation into an antitriplet.  This 
structure automatically cancels all anomalies, and when combined with 
the requirement of asymptotic freedom, necessitates that  the 
number of generations is equal to three.  The model has 
an automatic Peccei-Quinn symmetry\cite{pq,dias}.  The fact that one 
of the quark families is treated differently than the other two could 
lead to an explanation of the heavy top quark mass\cite{fram}.  This 
minimal model contains doubly charged bilepton gauge fields, as well 
as isosinglet quarks with exotic charges, leading to a rich 
phenomenolgy\cite{lots}.  A particularly exciting feature of this 
model is that there is an {\it upper} bound on the scale of 
$SU(3)_{L}$ breaking which is within range of the LHC.

In another version of the model, with a different embedding of the charge 
operator into $SU(3)_{L}\times U(1)$, the charged lepton in the 
antitriplet is replaced by a right-handed 
neutrino\cite{montero,foot}.  In this version, the bileptons are singly 
charged or neutral.   Another model can be 
found in which there are no lepton-number violating gauge bosons and 
no exotic quark charges (at the price of adding an isosinglet charged lepton 
for each generation).  Nonetheless, in all of these models, one still treats one of the quark 
generations differently than the other two.

It is most natural to have the third generation be the ``different'' 
generation, since this might explain the heavy top quark and since 
some of the constraints to be discussed below are substantially weakened.  With 
generations treated differently, one will expect to have tree-level 
flavor-changing neutral currents (FCNC).  Thus, it is expected that 
FCNC involving the third generation will be dominant.  Given the success 
of BELLE and BABAR, an analysis (and update of previous analyses) of rare B decays and FCNC in these 
models seems warranted.

In the next section, we discuss the three models mentioned above, as 
well as two other models in which all of the generations are treated 
identically.  In section III, we analyze current bounds from FCNC 
processes in these models.  Section IV contains our conclusions.

\section{Models}

A comprehensive review of the gauge, fermion and scalar sectors of the 
various
$SU(3)_{L}\times U(1)$ models can be found in Refs. \cite{ponce1} 
and \cite{ponce2}.  In this section, we briefly summarize this 
review, and then turn to a discussion of FCNC and rare B decays in 
these models.   Different models 
can be distinguished by the embedding of the electric charge 
operator.  In general, the charge operator is given by 
\begin{equation}
    Q = a T_{3L} + {2\over\sqrt{3}} b T_{8L} + x I_{3},
    \end{equation}
where we have used conventional normalization ($T_{i}=\lambda_{i}/2$ 
and Tr$(\lambda_{i}\lambda_{j}) = 2 \delta_{ij}$), $I_{3}$ is the 3x3 
unit matrix, and $a$ and $b$ are arbitrary.  The value of $x$ can be 
absorbed into the hypercharge definition, and will not be relevant.  
The fact that weak isospin is contained within the $SU(3)_{L}$ group 
implies that $a=1$, and so models are distinguished by the value of $b$.
It should be noted that gauge bosons will have integral charge only 
for half-integral values of $b$, and that models with negative $b$ can 
be transformed into models with positive $b$ by replacing triplet 
fermion representations with antitriplets, and vice versa.

The two choices for $b$ that have been considered are $b=3/2$ and 
$b=1/2$.   The former gives the original, minimal 
Pisano-Pleitez-Frampton model, with exotic isosinglet quark charges, 
while the latter does not lead to any exotic quark charges.  We now 
discuss each choice.

Of the 9 gauge bosons of the electroweak group, 3 are neutral and 
there are three charged pairs, the usual $W^{\pm}$ and two others with 
charges $\pm(b+1/2)$ and $\pm(b-1/2)$.  Thus this $b=3/2$ model has doubly charged 
gauge bosons.  In the minimal model, the fermion representations are
\begin{equation}
    L_{i} = \pmatrix{e\cr\nu_{e}\cr e^{c}\cr}, 
    \pmatrix{\mu\cr\nu_{\mu}\cr 
    \mu^{c}\cr},\pmatrix{\tau\cr\nu_{\tau}\cr \tau^{c}\cr}
    \end{equation}
for the leptons, and 
\begin{equation}
    Q_{i}=\pmatrix{u\cr d\cr D\cr},\pmatrix{c\cr s\cr 
    S\cr},\pmatrix{b\cr t\cr T\cr}\end{equation}
for the left-handed quarks.  The conjugates for these nine fields are 
all $SU(3)_{L}$ singlets.  $D$ and $S$ are new isosinglet quarks with 
charge $-4/3$ and $T$ is an isosinglet quark with charge $5/3$.  Note 
how the third generation is treated very differently than the first 
two.  This is necessary to cancel anomalies.  In principle, either of the three generations could be chosen 
to be different, however, as will be seen shortly, the strong bounds on FCNC in the kaon sector 
make it more likely that the third generation is singled out.  This 
is the original, minimal model, and will be referred to as Model A.

When an extension of the standard model predicts new phenomena, one 
can often explain non-observation of the phenomena by increasing the 
mass scale of the new physics.   However, that is not possible for 
the minimal $SU(3)_{L}\times U(1)$ model.   The reason is that if one were to 
embed the standard model entirely into the $SU(3)_{L}$ group, then
the unification gives $\sin^{2}\theta_{W}=1/4$.  
The extra $U(1)$ factor then forces one to have $\sin^{2}\theta_{W} 
\le 1/4$.   This is, of course, valid at low energy, but since 
$\sin^{2}\theta_{W}$ increases with scale, the scale of $SU(3)_{L}$ 
breaking cannot be too high.  In the original, minimal model, model A, 
this scale was estimated to be approximately $800$ GeV.  It has been 
argued\cite{findthis} that more precise definitions of ``scale'' 
allow this upper bound to be somewhat higher, as high as 2-3 TeV.
Thus, the model is capable of being ruled out in the near future.

A simple alternative to this model\cite{tully} is to change the lepton structure 
by replacing the $e^{c}_{i}$ with a heavy lepton $E^{+}_{i}$ and 
adding $e^{c}_{i}$ and $E^{-}_{i}$ singlets.   This will be 
referred to as Model $A^{\prime}$.

Although one can, of course, add a right-handed neutrino singlet to 
the above structure, the model of Montero, et al.\cite{montero,foot} 
modifies the lepton sector, and has, with $b=-1/2,$
\begin{equation}
    L_{i}=\pmatrix{\nu_{i}\cr e_{i}\cr \nu_{i}^{c}\cr}
    \end{equation}
with the $e_{i}^{c}$ being an $SU(3)_{L}$ singlet.  The quarks are 
given by
\begin{equation}
    Q_{i}=\pmatrix{d\cr u\cr D\cr},\pmatrix{s\cr c\cr 
    S\cr},\pmatrix{t\cr b\cr T\cr}\end{equation}
The new weak isosinglet quarks now have the same charges as their 
standard model counterparts, and the bileptons are either neutral or 
singly-charged.  This model will be referred to as Model B.

Since additional exotic quarks must be introduced in these models, it 
is natural, in the spirit of grand unification, to suppose that 
additional charged leptons are present.  In Model C, the leptons are 
taken to be 
\begin{equation}
    L_{i} = \pmatrix{\nu_{i}\cr e_{i}\cr E_{i}\cr}
    \end{equation}
and the quarks are
\begin{equation}
    Q_{i}=\pmatrix{d\cr u\cr U\cr},\pmatrix{s\cr c\cr 
    C\cr},\pmatrix{t\cr b\cr B\cr}
    \end{equation}
with all other fields (including right handed neutrinos, if necessary) 
being $SU(3)_{L}$ singlets.  This model has been explored in Ref. 
\cite{ozer}    

In all of the above models, the quark generations are treated 
differently.  There are two other models\cite{ponce1,ponce2} with identical quark 
generations to the previous two models, but in which the leptons are 
all treated very differently.  These models have not been explored in 
detail, and since we are interested in FCNC in the quark sector, they 
will not be discussed further here.

Finally, there are two models in which all generations, quark and 
leptons are treated equally.  These models lose the appealing feature 
of explaining the number of generations (via anomaly cancellation), 
but do have the feature of following naturally from grand unified 
theories.  In each of these models, there are $27$ fields in each 
generation.  In Model D, these fields fill out a $27$ of $E_{6}$, and 
arises naturally from the $E_{6}$ GUT.    This model 
has been analyzed in Ref. \cite{sanchez}.   Model E has a ``flipped'' 
structure, and  arises from an 
$SU(6)\times U(1)$ unified gauge symmetry, and has been discussed in 
Ref. \cite{mart}.    

Nothing in the above discussion is new, and there has been some 
phenomenological work on all of these models.  However, there has 
been very little done (especially in the three-generation models A,B 
and C) regarding FCNC B-decays, and the bounds from $\Delta m_{B}$ 
and $\Delta M_{B_{s}}$ need to be updated.  We turn to these issues in the next section.

It should be pointed out that the scalar sector of these models 
all contain at least three $SU(3)_{L}$ triplets\cite{higgs}, and in 
some cases an additional Higgs sextet is needed to give leptons 
mass\cite{diaz}.  
These Higgs triplets may give additional contributions to FCNC 
processes.  However, since these contributions will depend on large 
numbers of arbitrary parameters, we will ignore them--their inclusion 
would only strengthen the lower bounds on gauge boson masses (unless 
they interfere destructively and one fine-tunes).

\section{FCNC and rare B decays}

With different generations treated differently, it is not surprising 
that tree level flavor-changing neutral currents will arise.  
A nice discussion of FCNC interactions in the minimal model, model A, 
can be found in the works of Liu\cite{liu} and Gomez Dumm, et 
al\cite{dumm}.  They show that
\begin{equation}
    {\cal L}_{FCNC}={g\over \cos\theta_{W}}{1\over 
    2\sqrt{3}\sqrt{1-4\sin^{2}\theta_{W}}}(-\sin\phi Z_{1\mu}+\cos\phi 
    Z_{2\mu})J^{\mu}_{FCNC}\end{equation}
where $\phi$ is the mixing angle between the weak eigenstate $Z$'s and 
the mass eigenstates.  Since electroweak precision fits force this 
angle to be very small\cite{ng}, we will not include it (although will discuss 
possible interference terms later). Thus, $Z_{2}$ is approximately 
$Z^{\prime}$.  Note the fact that if 
$\sin^{2}\theta_{W}$ is greater than $1/4$, this breaks down, as 
discussed above.   The current is
\begin{equation}  
J^{\mu}_{FCNC}=2\cos^{2}\theta_{W}\overline{q}\gamma^{\mu}P_{L}q
\end{equation}
where $P_{L}$ is the left-handed projection operator.  In terms 
of mass eigenstates, this gives 
\begin{equation}
J^{\mu}_{FCNC} = 2\cos^{2}\theta_{W}\left( 
\overline{u}\gamma^{\mu}P_{L} U^{\dagger}_{L}\pmatrix{0&0&0\cr 
0&0&0\cr 0&0&1\cr} U_{L}u +
\overline{d}\gamma^{\mu}P_{L} V^{\dagger}_{L}\pmatrix{0&0&0\cr 
0&0&0\cr 0&0&1\cr} V_{L}d
\right)\end{equation}
where $U_{L}$ and $V_{L}$ diagonalize the left-handed $Q=2/3$ and 
$Q=-1/3$ quark mass matrices, respectively.

The $U_{L}$ and $V_{L}$ matrices are not independent, since one 
knows that $V_{CKM}=U_{L}^{\dagger}V_{L}$, but the individual values 
are not known.  FCNC processes will then depend on either $U_{L}$ or 
$V_{L}$ matrices alone, and one will not, without further assumptions, 
know their values.

The papers of Liu\cite{liu} and Gomez Dumm et al.\cite{dumm} 
calculate the $P-\overline{P}$ mass difference in this model.  For $\Delta m_{K}$, 
for example, they find that 
\begin{equation}
\Delta m_{K}={2\sqrt{2}\over 9}G_{F}{\cos^{4}\theta_{W}\over 
1-4\sin^{2}\theta_{W}}\big|V_{31}^{*}V_{32}\big|^{2}\eta_{Z}B_{K}f^{2}_{K}m_{K}
\left({M^{2}_{Z}\over m^{2}_{Z^{\prime}}}\right)\end{equation}
Here, $\eta_{Z}$ is a QCD correction factor, $B_{K}$ and $f_{K}$ are 
the bag constant and kaon decay constant.  Similar expressions can 
be obtained for other pseudoscalar systems.

Since there is an uncertainty of roughly a factor of two in the 
Standard Model expression, we assume that the contribution
for $K-\overline{K}$ is less than the Standard 
Model value, and that the $D-\overline{D}$ mixing is less than its 
experimental limit.  (In previous works, similar assumptions were 
made for the B systems.)  For $B-\overline{B}$ mixing, there is very little 
uncertainty in the hadronic matrix elements, and the primary 
uncertainty comes from $B_{B}$ and $f_{B}$, which give an uncertainty 
of approximately $30\%$; we assume the contribution is less than this 
uncertainty.  For $B_{s}-\overline{B}_{s}$ mixing, we require that the 
contribution be less than 10 picoseconds (for the oscillation time), 
since that is roughly the current uncertainty.  Using updated experimental values, 
we find the bounds in the first column of Table 1.

One can use these results, as done by Liu\cite{liu} to bound the 
mixing angles.   Alternatively, one can assume a Fritzsch-like 
structure\cite{dumm}, and write (with $i\geq j$) $V_{ij}=\sqrt{m_{j}/m_{i}}$ (similarly for 
$U_{ij}$), and then find bounds on $m_{Z^{\prime}}$.   Doing so gives 
an upper bound on $m_{Z^{\prime}}$, in TeV units, shown also in the 
first column of Table 1.  These bounds, especially for the $B-\overline{B}$ system, 
are very severe, and are well in excess of the upper bound on the 
$Z^{\prime}$ mass.  The angles must thus be smaller than one's naive 
expectation, or the model is excluded.  It is also shown by 
Liu\cite{liu} and Gomez Dumm\cite{dumm} that if one chose the first 
or second generation fields to be picked out as being different, then 
the bound would be much, much stronger, closer to $1000$ TeV.

The success of the B-factories has led to stringent bounds on 
$B\rightarrow K f^{+}f^{-}$, $B\rightarrow f^{+}f^{-}$ and 
$B_{s}\rightarrow f^{+}f^{-}$.  We now calculate these processes in 
this model. 

For $B\rightarrow K f^{+}f^{-}$, only the vector part of the 
interaction will contribute, and thus the matrix element $\langle 
K|\overline{s}\gamma^{\mu}b|B\rangle$ is needed.  We use the matrix 
elements of Isgur, et al. \cite{isgur}, as discussed in Ref. 
\cite{black}, which gives a value of $2f_{+}p_{K}^{\mu}$, where 
$f_{+}$ is given by ${3\sqrt{2}\over 8}\sqrt{m_{b}\over 
m_{q}}\exp({m_{K}-E_{K}\over m_{K}})$.  Here, $m_{q}$ is taken to be a 
constituent quark mass, or $300$ MeV.  Given this matrix element, the 
calculation is straightforward, and we find that the partial width is 
given, in GeV units,  by $\Gamma = 1.7\times 
10^{-15}V_{32}^{2}\left({M_{Z}\over M_{Z^{\prime}}}\right)^{4}$  Using 
the experimental bound and the Fritzsch ansatz, we find a bound of 
$1.2$ TeV on the mass of the $Z^{\prime}$, as seen in Table 1.
This is substantially weaker than the 
bound from $B_{s}-\overline{B}_{s}$ mixing.

For $B_{s}\rightarrow f^{+}f^{-}$, only the axial vector part of the 
interaction contributes.   Note that a helicity suppression makes the 
branching ratio proportional to the square of the final state fermion 
mass.  The best experimental bounds are for muon final states 
($B_{s}\rightarrow\tau^{+}\tau^{-}$ would be very interesting if one 
could come within a factor of a few hundred of the muonic branching 
ratio).  The standard axial vector matrix element $\langle 0 | 
\overline{s}\gamma^{\mu}\gamma_{5}b|B_{s}\rangle = f_{B_{s}} p^{\mu}$ 
is used, and we find that 
\begin{equation}
    \Gamma = {G^{2}_{F}M^{4}_{Z}f_{B}^{2}V_{32}^{2}m_{B}m^{2}_{\mu}\over 
    36\pi M^{4}_{Z^{\prime}}}\end{equation}
    Comparing with the experimental bound and using the Fritzsch 
    ansatz gives a lower bound of $0.23$ TeV on the $Z^{\prime}$ mass.
For $B\rightarrow f^{+}f^{-}$, we find very similar numerical 
results.  Again, this is substantially weaker than the bound from 
mixing.

\begin{table}[tbh]
{\large
\centering
\begin{tabular}{||c|c|c|c|c||}
\hline \hline
 & Model $A$ & Model $A'$ & Model $B$ & Model $C$ \\ \hline\hline
$\Delta m_K$ & ${1.6 \times 10^{-4}}$ & ${1.6 \times 10^{-4}}$ & $4.7 \times 10^{-4}$ & $1.7 \times 10^{-4}$ \\  
 & 4.8 TeV & 4.8 TeV & 1.7 TeV & 4.5 TeV \\ \hline
$\Delta m_D$ & ${1.6 \times 10^{-4}}$ & ${1.6 \times 10^{-4}}$ & ${4.8 \times 10^{-4}}$ & ${1.8 \times 10^{-4}}$ \\ 
& 250 GeV & 250 GeV & 80 GeV & 220 GeV \\ \hline
$\Delta m_B$ & ${1.4 \times 10^{-4}}$ & ${1.4 \times 10^{-4}}$ & ${4.1 \times 10^{-4}}$ & ${1.5 \times 10^{-4}}$ \\
& 30.7 TeV & 30.7 TeV & 10.5 TeV & 28.2 TeV \\ \hline
$\Delta m_{B_s}$ & ${1.1 \times 10^{-3}}$ &${1.1 \times 10^{-3}}$ & ${3.3 \times 10^{-3}}$ & ${1.2 \times 10^{-3}}$ \\ 
& 14.7 TeV & 14.7 TeV & 5.0 TeV & 13.5 TeV \\ \hline
$B_{d,s} \to \mu^+ \mu^-$& $0.15$ & $0.038$ & ${0.11}$ & ${0.32}$ \\  
& 230 GeV & 1.0 TeV & 340 GeV & 121 GeV \\ \hline
$B \to K \mu^+ \mu^-$ & ${3.2 \times 10^{-2}}$ & ${9 \times 10^{-3}}$ & ${3.5 \times 10^{-2}}$ & ${4.6 \times 10^{-2}}$ \\  
& 1.2 TeV & 4.3 TeV & 1.1 TeV & 800 GeV \\ \hline \hline
\end{tabular}
}
	\caption{Bounds on the models described in the text from several 
	flavor changing neutral processes.  The upper number is the bound 
	on $|V_{3i}^{*}V_{3j}|{m_{Z}\over m_{Z^{\prime}}}$, where $i$ and 
	$j$ refer to the relevant quark masses (and the V's are replaced by 
	$U$'s for $\Delta m_{D}$); for the rare B decays, the upper number 
	is the bound on $|V_{3i}^{*}V_{3j}|^{1/2}{m_{Z}\over m_{Z^{\prime}}}$.  
	The lower number is the lower bound on 
	the $Z^{\prime}$ mass assuming a Fritzsch structure for the $V$ 
	matrix.} 
\label{Table}     

\end{table}

It is important to note that even if one abandoned the Fritzsch ansatz (as one 
must for the model to be phenomenologically acceptable), the bound 
from quark-antiquark mixing will always be stronger (unless $V_{32}$ 
is exceptionally small (less than $10^{-3}$), in which case the bound 
on $m_{Z^{\prime}}$ is less than the direct search bound).  In short, there 
can be {\it no substantial contribution} to these rare B-decays in this 
model (since a substantial contribution would lead to an overly large 
contribution to $B-\overline{B}$ mixing), and this statement is 
independent of the mixing angles.   It should also be noted that we 
have ignored contributions from $Z$-exchange and from flavor-changing 
neutral Higgs exchange.  These could destructively interfere, 
weakening the bounds.  However, this would require some fine-tuning 
and since the Higgs sector has many free parameters, we do not 
consider this possibility.

In model $A^{\prime}$, the only difference is in the coupling of the 
final state leptons to the $Z^{\prime}$. While the mass differences 
are unchanged, there are substantial changes in rare B decays.  We 
find the bounds (see Table 1) on $B\rightarrow 
K \mu^{+}\mu^{-}$ to be $4.3$ TeV, and the bound from 
$B_{s}\rightarrow \mu^{+}\mu^{-}$ to be $1.0$ TeV.  Again, the bounds 
from the mass difference in the $B-\overline{B}$ system are stronger.
    
We now turn to the $b=1/2$ models.  The embedding of the charge 
operator now no longer forces $\sin^{2}\theta_{W}$ to be less than 
$1/4$, and thus the upper bound on the scale of $SU(3)_{L}$ breaking 
no longer applies.  As a result, the factors of 
$1-4\sin^{2}\theta_{W}$ end up being replaced by 
$1-{4\over 3}\sin^{2}\theta_{W}$.   

In Model B, the mass differences in the neutral $K$, $D$ and $B$ 
system (but not the $B_{s}$) were calculated in Ref. \cite{longvan}, 
and the bounds from the rare kaon decay $K^{+}\rightarrow 
\pi^{+}\nu\overline{\nu}$ were calculated\cite{rarek}.  We have reanalyzed these 
bounds, using updated constraints, and included the bounds from  the 
$B_{s}$ mass difference, and the rare $B$ and $B_{s}$ decays discussed 
above.

Again, if one assumes a Fritzsch-type structure for the $U$ and $V$ 
matrices, lower bounds on the $Z^{\prime}$ mass are obtained (one can 
easily remove that assumption and present results in terms of, for 
example, the $V_{ij}$ and quark masses).  The calculation is the same 
as for Model A, with different couplings.  We find the bounds listed 
in the third column of Table 1.  Again, the bounds from the mass differences are 
much stronger than from rare B-decays, and are weaker than for Model 
A (primarily due to the absence of a $1-4\sin^{2}\theta_{W}$ factor).

In Model C, the only calculation of flavor-changing neutral current 
effects that we are aware of is the calculation of the mass difference 
in the neutral kaon system by Ozer, in Ref. \cite{ozer}.  The fourth 
column of Table 1 lists these bounds.    The bounds from mass differences are 
substantially stronger than in model B.

Models D and E are very different.  They are one family models, and 
thus all generations are treated identically.  Due to the existence of 
isosinglet quarks, there will be flavor changing neutral currents.   
These models are explicitly explored in Refs. \cite{sanchez} and \cite{mart}.
FCNC in models with isosinglets have been explored in great detail in a 
number of papers.  The most recent is by Andre and 
Rosner\cite{rosner}; the reader is referred to that work and 
references therein.   In most of these works, it is assumed that there 
is only a singlet isosinglet quark (or if there are more than one, it 
is assumed that one is much lighter and thus dominates the physical 
effects), and thus the $Q=-1/3$ mass matrix 
is $4\times 4$, and it is often assumed that the $V_{34}$ element is 
the largest.  However, the models D and E contain three isosinglet 
quarks, and if the mass hierarchy of these quarks follows the 
standard mass hierarchy, the lightest of these will interact much more 
strongly with the down quark, i.e. the biggest element will be 
$V_{14}$.  An analysis of the phenomenology of this case would be 
interesting.

\section{Conclusions}

$SU(3)_{L}\times U(1)$ models fall into two categories, depending on 
the embedding of the charge operator into the $SU(3)_{L}$ group.  The 
choices of fermion representations further subdivides the models.  
These models all have tree-level FCNC mediated by gauge bosons.  We 
have calculated the $P-\overline{P}$ mass
differences and several rare B decays in these models.  In all cases, 
we find that the contribution from rare B decays is much smaller than 
those from $B-\overline{B}$ and $B_{s}-\overline{B}_{s}$ mass 
differences, and thus the models explicitly predict that there will 
be no substantial contribution to these rare B-decays (independent of 
mixing angles).   Lower bounds on gauge boson masses are typically 
of the order of tens of TeV if one assumes a Fritzch-like structure 
for the mixing angles.   This is a serious problem for the original, 
minimal model, which has an upper bound of approximately 2-3 TeV for 
the gauge boson masses.  Thus, these models can only survive if the 
mixing angles are much smaller than one's naive expectation.  This 
would mean that the down-quark mixing matrix would be very nearly 
diagonal, and thus CKM mixing would have to arise from the $Q=2/3$ 
sector.   This severely constrains attempts to understand the origin 
of flavor in these models.

We thank Andrzej Buras for useful discussions.  JR would like to 
thank Colciencias and DIB for financial support and the College of William and 
Mary for its hospitality.  The work of MS was supported by the 
National Science Foundation grant PHY-023400.

\end{document}